\documentclass[10pt,twocolumn,letterpaper]{article}

\usepackage{cvpr}              % To produce the CAMERA-READY version
\definecolor{cvprblue}{rgb}{0.21,0.49,0.74}
\usepackage[pagebackref,breaklinks,colorlinks,allcolors=cvprblue]{hyperref}
\usepackage{makecell}

\def\paperID{16702} % *** Enter the Paper ID here
\def\confName{CVPR}
\def\confYear{2025}

\title{Exploring adversarial robustness of JPEG AI: methodology, comparison and new methods}

\author {
    Egor Kovalev\textsuperscript{\rm 3},
    Georgii Bychkov\textsuperscript{\rm 2,3},
    Khaled Abud\textsuperscript{\rm 3},
    Aleksandr Gushchin\textsuperscript{\rm 1,2,3},
    Anna Chistyakova\textsuperscript{\rm 2,3},\\
    Sergey Lavrushkin\textsuperscript{\rm 1,2},
    Dmitriy Vatolin\textsuperscript{\rm 1,2,3}
    Anastasia Antsiferova\textsuperscript{\rm 1,2,4}\\
    \textsuperscript{\rm 1}MSU Institute for Artificial Intelligence\\
    \textsuperscript{\rm 2}ISP RAS Research Center for Trusted Artificial Intelligence\\
    \textsuperscript{\rm 3}Lomonosov Moscow State University\\
    \textsuperscript{\rm 4}Laboratory of Innovative Technologies for Processing Video Content, Innopolis University\\
    \{egor.kovalev, georgii.bychkov, khaled.abud, alexander.gushchin, anna.chistyakova, \\sergey.lavrushkin, dmitriy, aantsiferova\}@graphics.cs.msu.ru
}

\begin{document}
\maketitle
\begin{abstract}
Adversarial robustness of neural networks is an increasingly important area of research, combining studies on computer vision models, large language models (LLMs), and others. With the release of JPEG AI --- the first standard for end-to-end neural image compression (NIC) methods --- the question of its robustness has become critically significant. JPEG AI is among the first international, real-world applications of neural-network-based models to be embedded in consumer devices. However, research on NIC robustness has been limited to open-source codecs and a narrow range of attacks. This paper proposes a new methodology for measuring NIC robustness to adversarial attacks. We present the first large-scale evaluation of JPEG AI's robustness, comparing it with other NIC models. Our evaluation results and code are publicly available online (link is hidden for a blind review).
\end{abstract}    
\section{Introduction}
With the success of neural networks in various high-level computer vision tasks, the researchers proposed many low-level image processing methods, such as image compression, based on neural networks~\cite{nic1, nic2, nic3}. These methods use neural networks to solve sub-tasks of conventional image compression pipelines or propose an end-to-end pipeline for the encoding-decoding process. Promising results shown by NIC methods led to the establishment of JPEG AI \cite{jpeg_ai_standard}, the first complete image compression standard based on neural networks. It is a Joint Photographic Experts Group (JPEG) project that promises significant improvements over non-neural-networks-based codecs, such as JPEG or JPEG2000. Image compression in JPEG AI is trained end-to-end and includes four key components: analysis transform, where an encoder transforms the original image into a latent representation using a nonlinear transform; quantization, used to compress the image representation and reduce the excessive information; entropy coding, where data is compressed into binary data based on the predictability of the information; and synthesis transform, where using a learned nonlinear synthesis transform, a decoder reconstructs the image from the compressed data. Standardization involves developing a common testing methodology, including test sequences, quality measurement methods, etc.  

The adversarial robustness of neural networks has been a growing area of research recently \cite{attacks1, chakraborty2021survey, zhang2021survey, wei2024physical}. Numerous studies have been published regarding adversarial attacks, defenses, and overall analysis of robustness for various computer vision tasks, such as image classification \cite{Zhu2024attack}, detection \cite{nezami2021pick}, quality assessment \cite{iqarobustness}, etc.
\begin{figure*}[htb]
  \centering
   \includegraphics[width=\linewidth]{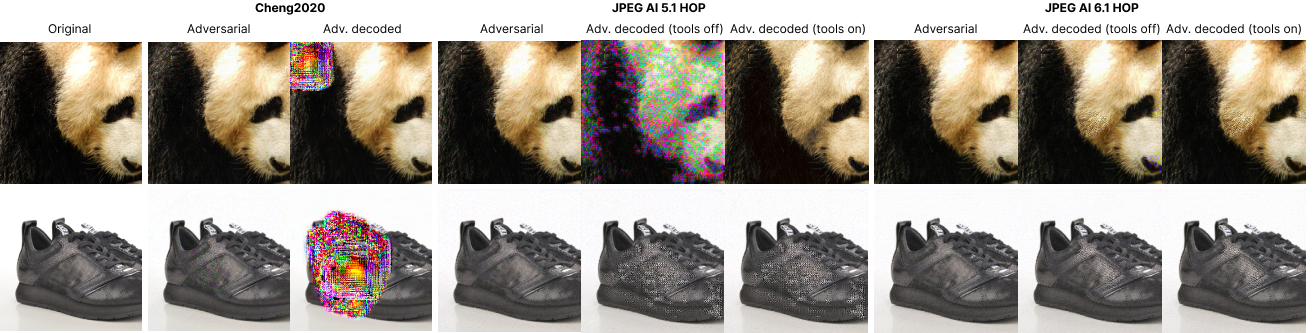}
  \caption{Examples of adversarial attacks on neural image compression methods and comparison of JPEG AI versions under attack. The attack is constructed by the MADC method.}
  \label{fig:example}
\end{figure*}
With the release of a new promising algorithm for compression that may become a standard approach in the industry, it is crucial to assess its robustness against adversarial threats to mitigate potential risks in image compression systems. To accomplish this, we thoroughly analyzed versions of JPEG AI and its latest counterparts by subjecting them to adversarial attacks with various loss functions. We also explore how to counter adversarial attacks with preprocessing defense strategies. Our main contributions are summarized as follows:
\begin{itemize}
    \item We extend the methodology proposed in \cite{ftda} by incorporating multiple full-reference metrics—namely, $\Delta \text{PSNR}$, $\Delta \text{MSE}$, $\Delta\text{MS-SSIM} $, and $\Delta \text{VMAF}$ --- to evaluate robustness under attack.  %Additionally, we introduce $D_{score}$ to measure the distance between the benign image and its adversarial counterpart. These adjustments allow us to capture compressing results more comprehensively.
    \item We perform extensive experiments on 10 NIC models (including novel JPEG AI), analyzing 6 attacks targeting image quality or bitrate. To the best of our knowledge, it is the first study that evaluates NICs across diverse attack-loss combinations, providing insights into the relative effectiveness of attacks, which were performed with 6 different loss functions. We also analyze how attacks affect computer vision models after neural-based compression.  
    \item We further investigate how to overcome vulnerabilities in NICs by applying purification defense strategies. Our work compares not only codecs themselves but also attacks and defenses. 
\end{itemize}
%The source code for reproduction can be found here: \textit{link hidden}.

\section{Related work}
\subsection{Neural image compression}

Neural image compression has been developing rapidly in recent years. Balle \etal \cite{balle2016end} use a generalized divisive normalization of joint nonlinearity and uniform scalar quantization for image compression. \cite{agustsson2017soft} implemented a new soft-to-hard vector quantization method and incorporated it into the compressive autoencoder introduced earlier. Balle \etal \cite{balle2018variational} express the image compression problem as variational autoencoders and propose hyperpriors to improve the entropy model. Minnen \etal \cite{minnen2018joint} improve the GSM-based entropy model by generalizing the hierarchical GSM model to a Gaussian mixture model and adding an autoregressive component. Mentzer \etal \cite{mentzer2020high} use GAN in their neural compression model, which helps to achieve high perceptual fidelity. Chang \etal \cite{cheng2020learned} develop a novel entropy model by leveraging discretized Gaussian mixture likelihoods combined with a simplified attention module to present an efficient image compression network. Yang \etal \cite{yang2021slimmable} presents a novel rate and complexity control mechanism with the help of slimmable modules. He \etal \cite{he2022elic} describe a new ELIC model that adopts stacked residual blocks as nonlinear transforms and uses the Space-Channel ConTeXt (SCCTX) model, which is a combination of the spatial context model and the channel conditional backward-adaptive entropy model. Zou \etal \cite{zou2022devil} introduce a flexible window-based attention module to enhance image compression models and train CNN and Transformer models that reach promising results. Liu \etal \cite{liu2023learned} propose using parallel transformer-CNN mixture blocks to combine the advantages of both approaches and a new entropy model that uses a swim-transformer-based attention module with channel squeezing. \cite{duan2023lossy} adopted a hierarchical VAE architecture, ResNet VAE, for image compression, using a uniform posterior and a Gaussian convolved with a uniform prior. Wang \etal \cite{wang2023evc} create a real-time neural image compression model using residual blocks and depth-wise convolution blocks and then use mask decay and novel sparsity regularization loss to transfer knowledge to smaller models. Yang \etal \cite{yang2024lossy} present a novel lossy compression scheme that uses an encoder to map images onto a contextual latent variable fed into a diffusion model for reconstructing the source images.

\subsection{JPEG AI and its versions}

JPEG AI \cite{jpeg_ai_standard} applies quantization across the entire image based on learned patterns, which is more efficient than the traditional block-based approach. The standard has a high operation point (HOP) and a base operation point (BOP). They differ in terms of compression efficiency and computational complexity. The HOP uses a multistage context modeling (MCM) technique and a more complex synthesis transform with an embedded attention mechanism. This allows for improved compression, but at the cost of reduced performance. The standard also includes a set of additional tools that can help make compression more efficient and adaptive to the content. These include Residual and Variation Scale (RVS), Filters such as Adaptive Re-Sampler, ICCI (Cross-Color Filter), LEF (Luma Edge Filter), and Non-Linear Chroma Filter, Latent Scale before Synthesis, and Channel-Wise Gain. The specific tools and their settings can vary depending on the codec configuration. Therefore, in the future, JPEG AI will be tested without specific tools, and the evaluation will be based on the default settings for each version of codec.

\subsection{Adversarial robustness of neural image compression}
Kang \etal \cite{malic} proposed an attack to increase the size of NIC-compressed images using the I-FGSM \cite{ifgsm} attack. Their study explored the attack's effectiveness across different codec architectures, identifying architectural elements that improved model reliability, and proposed their factorized attention model as the most stable. Chen and Ma \cite{chen2023toward} demonstrated the high vulnerability of NIC methods to adversarial attacks that reduce the quality of the decoded image. They adopted the I-FGSM and Carlini-Wagner \cite{cw} attacks to increase the difference between the compressed images before and after the attack. They attacked images after compression and proposed a new Fast Threshold-constrained Distortion Attack (FTDA). They also proposed a $\Delta \text{PSNR}$ method for assessing the success of an attack, which evaluates both its effectiveness and its visibility: 

\begin{equation*} 
\Delta \text{PSNR} = \text{PSNR}(x, C(x)) - \text{PSNR}(x', C(x')),
\end{equation*}
where $x$ is the original image, $x'$ is the adversarial image, $C(\cdot)$ - image after NIC.

%Our study builds on this approach by introducing the $D_{score}$ metric, an adaptation that aims to provide an enhanced perceptual measure of robustness.

% \begin{figure*}
%   \centering
%   \begin{subfigure}{0.68\linewidth}
%     \fbox{\rule{0pt}{2in} \rule{.9\linewidth}{0pt}}
%     \caption{An example of a subfigure.}
%     \label{fig:short-a}
%   \end{subfigure}
%   \hfill
%   \begin{subfigure}{0.28\linewidth}
%     \fbox{\rule{0pt}{2in} \rule{.9\linewidth}{0pt}}
%     \caption{Another example of a subfigure.}
%     \label{fig:short-b}
%   \end{subfigure}
%   \caption{Example of a short caption, which should be centered.}
%   \label{fig:short}
% \end{figure*}

\section{Problem statement}
\textbf{Neural image compression}.
Lossy image compression is based on a rate-distortion theory. The goal of lossy image compression is to find a trade-off between the size of the compressed image representation and the decrease in the perceptual quality of a reconstructed image. This problem is formulated as follows:
\begin{equation}
    \mathbb{E}_{x \sim p_x} [\lambda r(\hat{y}) + d(x,\hat{x})],
\end{equation}
where $r(\hat{y})$ is a bit representation of a quantized image after arithmetic encoding, and $d(x,\hat{x})$ --- perceptual similarity metric (PSNR, SSIM, etc.).

Auto-encoder architecture is one of the possible solutions. For a given image $x \in X=\mathbb{R}^{H\times W \times 3}$ of a distribution of natural images $p_x$, an encoder $E$ transforms it to a latent representation $y=E(x)$ of a distribution $p_y$. Then, the data is quantized $\hat{y}=Q(y)$, and a decoder $G$ performs the reconstruction of the image $\hat{x}=G(\hat{y})$. We denote $C(x)$ as a complete encoding-decoding process $C(\cdot) = G \circ Q \circ E: X \to X$.

\textbf{Adversarial attack on image compression.} The goal of an adversarial attack is to find a perturbation $\delta$ which, added to the original image, makes the adversarial image $x' = x + \delta$ such that its decoded image $C(x')$ differs from the original image as much as possible. Adversarial attack $A: X \rightarrow X$ is defined as follows:
\begin{equation}
    \begin{aligned}
    A(x)= \underset{x': \rho(x',x) \le \varepsilon} {\arg\max} \: L(x, x', C(x), C(x')),
    \end{aligned}
\end{equation}
where $\rho(x',x) = \|\delta\|$, $\varepsilon$ imposes a constraint on the perturbation magnitude, $L: X \times X \to \mathbb{R}$ is corresponding optimization target. To achieve this goal, we consider 10 loss functions for all employed attacks. They reflect different approaches to measuring the distance between the original and adversarial images and their reconstructed versions. Additionally, we consider an alternative optimization goal. Instead of increasing the distance between the image targets, we reduced the compression ratio of the NIC measured in Bits Per Pixel (BPP):
\begin{equation}
    \begin{aligned}
    A(x)= \underset{\delta: \|\delta\| \le \varepsilon} {\arg\max} \: \text{BPP}(Q(E(x + \delta))).
    \end{aligned}
\end{equation}
We list all optimization targets in Table \ref{table:losses}.

\begin{table}[h]
\centering
\begin{tabular}{lc}
Optimisation target & Formula \\ 
\midrule
FTDA default & $\left\| C(x) - C(x') \right\|_{2}$ \\
Added-noises & $\left\| C(x') - C(x) - (x' - x) \right\|_{2}$  \\
Reconstruction & $\left\| C(x') - x' \right\|_{2}$ \\
FTDA MS-SSIM & MS-SSIM$\left(C(x), C(x')\right)$ \\
\makecell[l]{Reconstruction\\MS-SSIM} & MS-SSIM$\left(x', C(x')\right)$ \\
BPP increase & $1 - bpp\left(C(x')\right)$ \\
\midrule
\makecell[l]{Y-modification\\of all targets} &  Take only Y in YCbCr \\ 
\end{tabular}
\caption{List of optimisation targets, where $x$ --- original image, $x'$ --- adversarial image, $C()$ --- image reconstruction after compression}
\label{table:losses}
% \end{floatrow}
\end{table}

\textbf{Adversarial purification.} Adversarial purification consists of two additional steps in the compression procedure: a preprocessing transformation $T: X \rightarrow X$ before compression and a postprocessing transformation $T^{-1}: X \rightarrow X$ afterward. %Adversarial purification methods aim to minimize the distortion between the reconstructed adversarial image and the original image. 
Mathematically, this can be expressed as:
% \begin{equation}
%     \begin{aligned}
% \min_{T} \; \mathbb{E}_{x, \delta} L(x, x + \delta, g(x), g(x + \delta)),
% \end{aligned}
% \end{equation}
% where $g(\cdot) = (T^{-1} \circ f \circ T)(\cdot)$ is the adversarially defended NIC.

\begin{equation}
\begin{aligned}
A(x) = \underset{x': \rho(x',x) \le \varepsilon} {\arg\max} \: L(x, x', g(x), g(x')),
\end{aligned}
\label{eq:defended_formula}
\end{equation}
where $x \in X$ is the original image, $x' \in X$ is the adversarial image, $g(\cdot) = (T^{-1} \circ C \circ T)(\cdot)$ is the adversarially defended NIC, $T(\cdot): X \to X$ is adversarial defense, $T^{-1}(\cdot): X \to X$ is the inverse of that defense. The equation means adding preprocessing step $T$ before feeding the image to NIC and postprocessing step $T^{-1}$ to reverse the preprocessing step. This equation can be transformed into an adversarial attack on undefended NIC when $T(X) = X$.

\section{Methodology}

\subsection{NIC models}
To evaluate the robustness of JPEG AI, we compare it with the robustness of other neural network image codecs. This further allows us to evaluate each NIC method's behavior and robustness. The codecs are listed in \autoref{table:list_of_codecs}.
% \begin{center}
\begin{table}[h]
% \footnotesize
\centering
\begin{tabular}{lccc}
Codec & Bitrates & Model type & Date \\ 
\midrule
\makecell[l]{JPEG AI 4.1 \\HOP/BOP} & 4 & \makecell{resid. \\ (context, \\attn. in hop)} & 2023 \\
\makecell[l]{JPEG AI 5.1 \\HOP/BOP} & 4 & \makecell{resid. \\ (context, \\attn. in hop)} & 2024 \\
\makecell[l]{JPEG AI 6.1 \\HOP/BOP} & 4 & \makecell{resid. \\ (context, \\attn. in hop)} & 2024 \\
\makecell[l]{Balle 2018 \\factorized,\\hyperprior \cite{balle2018variational}} & 4 & \makecell{factorized, \\hyperprior} & 2018 \\
Cdc \cite{yang2024lossy} & 3 & VAE+prior & 2024\\
\makecell[l]{Cheng2020,\\+attn. \cite{cheng2020learned}} & 4  & \makecell{resid., \\resid. + attn.} & 2020 \\
elic \cite{he2022elic} & 3 & context & 2022 \\
EVC \cite{evc} & 4 & context & 2021 \\
HiFiC \cite{mentzer2020high} & 3 & GAN-based & 2020 \\
Li-tcm \cite{liu2023learned} & 3 & --- & 2023 \\
\makecell[l]{mbt2018-6 bpp, \\mbt2018-mean\cite{minnen2018joint}} & 4 & context & 2018 \\
qres-vae \cite{duan2023lossy} & 3 & VAE+context & 2023 \\ 
\end{tabular}
\caption{List of NICs, with basic information about each one}
\label{table:list_of_codecs}
\end{table}
% \end{center}

\subsection{Attacks}
We use various adversarial attacks to assess a NIC model's robustness (Table \ref{table:attacks}). This allows us to create more diverse perturbations and draw more reasonable conclusions about the model's robustness.
\begin{table}[tb]
% \floatsetup{floatrowsep=qquad, captionskip=4pt}
% \begin{floatrow}

% \caption{Global caption}
\centering
\begin{tabular}{ll}
Attack & Description\\
\midrule
FTDA \cite{chen2023toward} & \makecell[l]{NIC attack to increase distance \\ between decoded images}\\
I-FGSM \cite{ifgsm} & Iterative sign gradient descent \\
MADC \cite{madc} & Proj. grad. on a proxy metric (MSE)\\
PGD \cite{pgd} & I-FGSM with random initialization \\
SSAH \cite{ssah} & Grad. desc. in high freq. domain \\
CAdv \cite{cadv} & Gradient descent with color filter \\
Random noise & Gaussian noise with $\sigma \in [\frac{5}{255}; \frac{14}{255}]$\\
\end{tabular}
\caption{List of evaluated attack methods and their authors.}
\label{table:attacks}
\end{table}
In this study, we focus on white-box attacks for the following reasons. Compression itself is a purification defense, and it can mitigate adversarial noise introduced by weaker black-box attacks, while white-box attacks offer more robust and effective perturbations. Also, black-box attacks are much more computationally expensive, severely limiting their range of applications. We choose six different white-box attacks of various types and Gaussian random noise to conclude that adversarial attacks are far more effective than random perturbations of the image. \textbf{MADC} \cite{madc} was one of the first methods introduced in 2008 that uses gradient projection onto a proxy Full-Reference metric to preserve image quality. \textbf{I-FGSM} \cite{ifgsm} is a well-known iterative modification of FGSM attack with simple sign gradient descent. \textbf{PGD} \cite{pgd} is similar to I-FGSM but uses random initialization. \textbf{SSAH} \cite{ssah} decompose an image into low- and high-frequency domains and insert perturbation in the latter to reduce attack visibility. \textbf{cAdv} \cite{cadv} attack applies a filter in LAB color space, shifting the color distribution of an image without introducing noise. We also included \textbf{Random noise} as a baseline attack. It samples Gaussian noise in an attempt to attack the model.

\subsection{Adversarial defenses}
\label{sec:defense_methodology}

\begin{table*}[h]
\centering
    \begin{tabular}{lccccl}
    Defense method & Type & Parameters & Preprocess ($Y = T(X)$) & Postprocess($X = T^{-1}(Y)$) \\
    \hline
    Flip & Spat. transf. & --- & $\text{flip}(X, [2, 3])$ & $\text{flip}(Y, [2, 3])$ \\
    Random roll & Spat. transf. & \makecell[c]{$\text{dim} \in \{2, 3\}$, \\ $\text{size} \in [0, \text{len}(X[\text{dim}]) - 1]$} & $\text{roll}(X, \text{size}, \text{dim})$& $\text{roll}(Y, -\text{size}, \text{dim})$\\
    Random rotate & Spat. transf. & $\theta \in [0, 359]$ & $\text{pad}(\text{rotate}(X, \theta))$& $\text{crop}(\text{rotate}(Y, -\theta))$  \\
    \makecell[l]{Random color \\reorder} & Color transf. & $\sigma: \{0, 1, 2\} \to \{0, 1, 2\}$ & $X[:, \sigma([0, 1, 2])]$& $Y[:, \sigma^{-1}([0, 1, 2])]$\\
    Random ens. & Ensemble & --- & Varies & Varies \\
    \makecell[l]{Geometric \\self-ens. \cite{chen2023toward}} & Ensemble & --- & Varies & Varies \\
    DiffPure \cite{nie2022diffusion} & Purification & --- & $\text{diffpure}(X)$ & $Y$ \\
    \end{tabular}
\caption{List of adversarial defenses used in our paper.}
\label{tab:defence_details}
\end{table*}

We selected several reversible adversarial defenses to evaluate their efficiency against adversarial attacks on NIC. We employed mainly reversible transformations to reduce their effect on image degradation. 
The full list of the applied defenses is in Table \ref{tab:defence_details}. \textbf{Flip} takes an image and reflects it horizontally or vertically. The reversed version flips the output image again to restore the original image orientation. \textbf{Random roll} selects either height or width randomly and samples the size of the roll, then rolls the image by a random number of pixels. The reverse step restores the original alignment. \textbf{Random rotate} defense samples the angle randomly and rotates the image on the selected angle. To make restoration of the original image possible, the method also performs a center pad of the original image to ensure image borders do not cut all its content. \textbf{Random color reorder} defense chooses perturbation of the color channels of the image tensor and swaps them, restoring the original order of the NIC output. \textbf{Random ensemble} combines Roll, Rotate, and Color reorder properties. It samples 10 actions from Roll, Rotate, and Color reorder with 4, 4, and 1 weights, respectively. \textbf{Geometric self-ensemble}~\cite{chen2023toward} generates 8 defended image candidates with flipping and rotation. It chooses one of them as the output that resembles the least distorted after the preprocess-NIC-postprocess pipeline. The distortion is measured by the mean squared error between the original image and the image processed by the defended NIC. \textbf{DiffPure}~\cite{nie2022diffusion} was developed to counter adversarial attacks on image classifiers and showed state-of-the-art performance for defending computer vision models. It performs purification based on a diffusion model as preprocessing and does nothing in the postprocessing step.

\subsection{Datasets}

% перечислить датасеты, написать почему они выбраны
% NIPS 100 images, KODAK 24 images, CityScapes 50 images, bsds
% TODO add about sampling
To evaluate methods, we chose four well-known datasets. The KODAK Photo CD \cite{kodakDataset} set consists of 24 $768 \times 512$ uncompressed images. CITYSCAPES \cite{cityscapesDataset} is a domain-specific dataset for image segmentation tasks with urban street scenes. These datasets are commonly used in the field of image compression. NIPS 2017: Adversarial Learning Development Set \cite{nipsDataset} is designed for evaluating adversarial attacks against image classifiers. Finally, the BSDS dataset \cite{bsdsDataset}, with 500 images of $320 \times 448$ resolution focusing on segmentation and boundary detection, enables us to evaluate the effects of NIC on segmentation model performance, further assessing implications of attacks on compression models for computer vision tasks.

\subsection{Quality metrics}
% \cite{ftda}
% \cite{nic robustness}
% delta fr metric (psnr, vmaf, msssim)
% arxiv версию статьи Ромы 
We employ four different Full-Reference image quality metrics to numerically assess the effects of adversarial attacks on images before and after the reconstruction: PSNR, MSE, MS-SSIM \cite{ms_ssim} and VMAF \cite{li2018vmaf}. PSNR, MSE, and MS-SSIM are traditional image-similarity measures. MS-SSIM \cite{ms_ssim} provides a scale-independent quality estimate, and VMAF implements a learning-based approach that aligns well with human perception \cite{NEURIPS2022_59ac9f01}. VMAF was designed to estimate the quality of distorted videos, but it can also be applied to images, interpreting them as single-frame videos.

Following the methodology of \cite{ftda}, we measure $\Delta_{score}$~--- the difference in reconstruction quality between the clear image and corresponding adversarial example:
\begin{equation}
 \Delta_{score}= FR(x, C(x)) - FR(x', C(x')),
\end{equation}
where $x$ is an original image, $x'$ is a corresponding adversarial example, $FR(x,y)$ is one of the aforementioned IQA models, and $C(x)$ is an evaluated NIC (entire encoding-decoding procedure). This score captures the drop in the fidelity of reconstructed images caused by the adversarial attack. Higher values of $\Delta_{score}$ indicate lower robustness of the NIC (if higher values of considered FR model correspond to better visual quality --- i.e., for PSNR, MS-SSIM, and VMAF).

To measure transferability of adversarial attacks to other codecs, we measure a modified version of $\Delta_{score}$:
\begin{equation}
\label{eq:delta_transf}
 \hat{\Delta}_{score}=FR(C(x_i), C(x_i')) - FR(C'(x_i), C'(x_i')),
\end{equation}
where $C(x)$ is the decoded image by the target codec used to construct an adversarial image, and $C'(x)$ is the attacked codec to which we test the attack's transferability. This metric offers a more precise evaluation for different pairs of models when comparing NIC with different robustness. 
%In addition to $\Delta_{score}$, we also use a slightly different metric to measure the effectiveness of the attacks on NICs:
%\begin{equation}
% D_{score}=FR(x, x^*) - FR(C(x), C(x^*)).
%\end{equation}
%It evaluates the difference between the distances between the benign image, its adversarial counterpart, and their reconstructed versions. For the ``higher—better'' FR model, high values of $D_{score}$ indicate that during the reconstruction of the adversarial example, NIC produced a result that diverges from the reconstructed benign image significantly stronger than the differences between input images. Therefore, higher average $D_{score}$ values correspond to lower robustness of the evaluated compression model.
%In section \ref{metrics_artifacts}, we compare these two evaluation scores ($\Delta_{score}$ and $D_{score}$) and conclude that they capture different concepts and do not show a high correlation with each other.

\subsection{Implementation details}

We used the source code of JPEG AI without additional pretraining. For adversarial attacks, to simplify backward propagation and avoid overfitting for extra tools, a modified interface for attacking the core model was used. As a result, constructed attacks depended less on the choice of the codec configuration. To evaluate the effectiveness of the attack and the codec's robustness, the attack's impact was assessed not only on the core model's result but also on the entire codec with additional tools included from the codec's base configuration. For this, we made only minor changes to the interface for working with the codec, according to our requirements, and used default settings provided by the authors. By implementing these changes that do not affect the codec's operation, we created a unified interface for three versions of JPEG AI: 4.1, 5.1 and 6.1, which allowed attacks to be trained and evaluated similarly to other neural network codecs presented in this paper.

We applied each adversarial attack to each encoder four times with varied attack parameters (learning rate, number of iterations, and perturbation bound). We then averaged the metrics for all launches.

Calculations were made with Slurm Workload Manager, 120 × NVIDIA Tesla A100 80 Gb GPU, and Intel Xeon Processor 354 (Ice Lake) 32-Core Processor @ 2.60 GHz. 
\section{Results}

\subsection{Comparison of adversarial attacks' efficiency}
\begin{figure*}[tb]
  \centering
\includegraphics[width=0.99\linewidth]{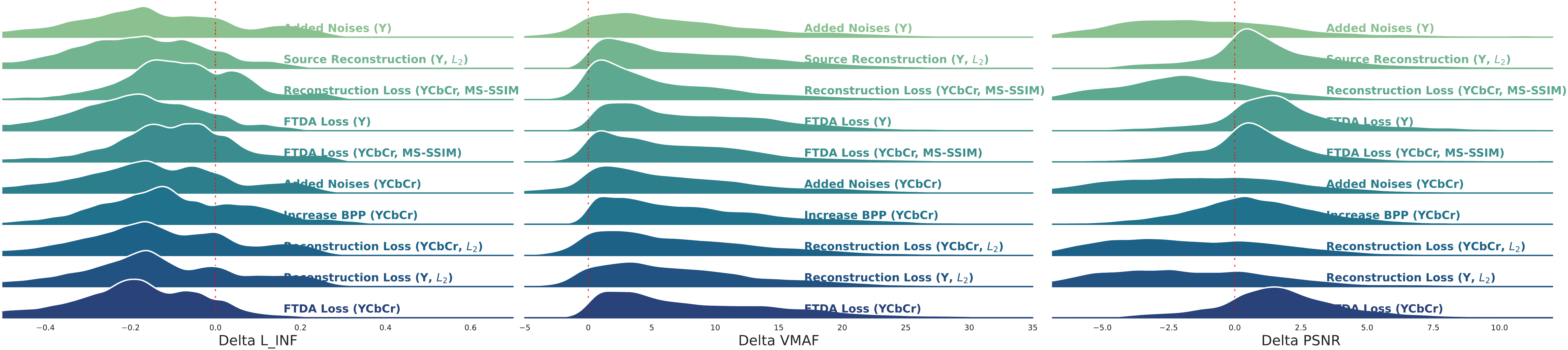}
  \caption{$\Delta$-metrics depending on the loss functions in all attacks.}
  \label{fig:losses}
\end{figure*}
In this section, we compare different loss functions used for adversarial attacks in terms of delta-quality metrics. Fig.~\ref{fig:losses} shows how they influence the quality of uncompressed images. Mostly, results across different $\Delta$ metrics are similar. The original loss proposed in the FTDA attack yields disturbance for most metrics, especially in terms of $\Delta L_{\infty}$. 
Reconstruction losses have similar results --- $\Delta \text{MS-SSIM}$, $\Delta \text{PSNR}$, and $\Delta \text{SSIM}$ generally decrease after attacks with these losses, meaning that the quality of decompressed images with attacks is better than without attacks. These losses directly maximize the difference between decoded and original images. More complex losses showed less efficiency.

\subsection{Comparison of codecs' robustness}
\begin{figure}[h]
  \centering
\includegraphics[width=\linewidth]{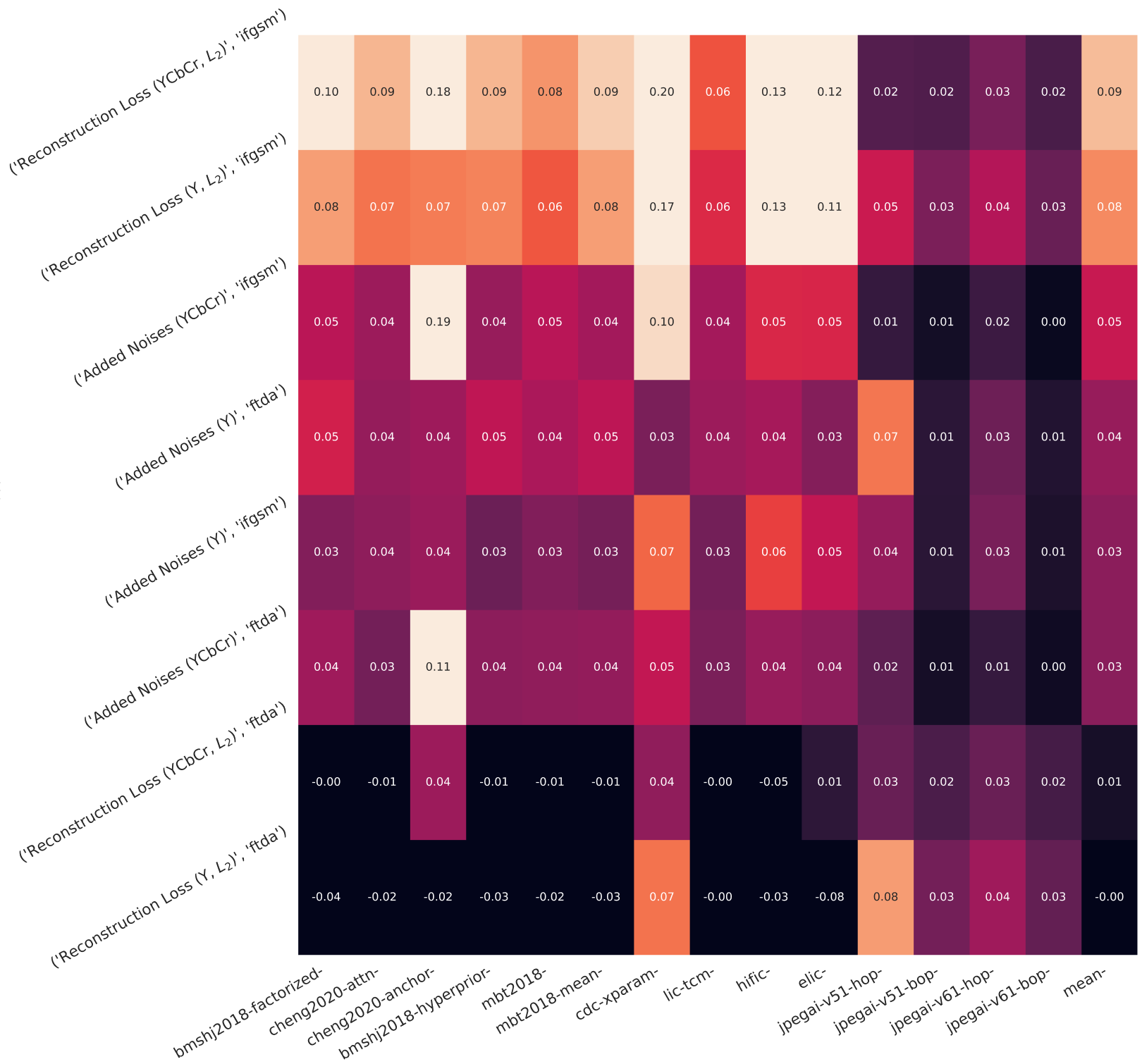}
   \caption{$\Delta$VMAF for analyzed NIC models under adversarial attacks with different loss functions. Dark colors refer to higher robustness.}
\label{fig:overall_robustness}
\end{figure}

Fig.~\ref{fig:overall_robustness} shows the adversarial robustness of tested NIC models. First of all, different codecs are vulnerable to different kinds of adversarial attacks. For example, Cheng2020 is subject to I-FGSM and FTDA attacks, which are ineffective against JPEG AI. JPEG AI showed relatively high robustness compared to other NIC models. High-operation point versions of JPEG AI are less robust than base-operation point. Second, the robustness of JPEG AI improved with a newer version (6.1 compared to 5.1). Also, JPEG AI is subject to attacks that operate only the luma component while creating the adversarial image (Y in loss function). Finally, the diffusion-based CDC method showed the lowest robustness to various attacks. This model may be less robust by design; adversarial noise causes significant changes in latent representation, yielding noticeable quality degradation.

\subsection{BPP increase}
\begin{figure*}[tb]
  \centering
\includegraphics[width=\linewidth]{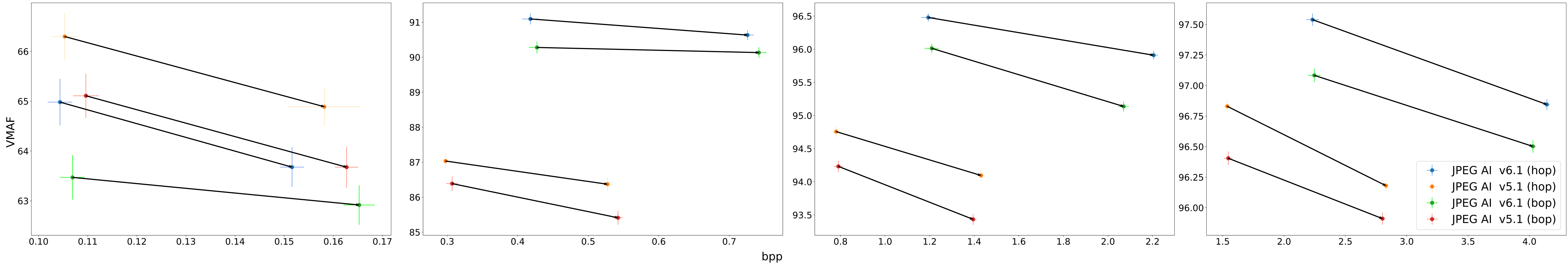}
  \caption{Change in compressed image's size before and after an attack for different target bitrates.}
  \label{fig:bpp}
\end{figure*}

We noticed that adversarial attacks increase the size of compressed images, even if an attack was not targeted to increase BPP.  Fig.~\ref{fig:bpp} shows this difference. This effect is explained by the more noise structure of adversarial images, which yields a different rade-distortion trade-off compared to benign images. Thus, adversarial images harm the reconstructed image's quality and the compressed representation's size.

\subsection{Codecs artifacts after attacks}
\label{metrics_artifacts}
To analyze artifacts introduced by adversarial attacks, we calculate two methods proposed in \cite{tsereh2024jpegaiimagecompression} to detect them. The first one, named ``Color metric'', analyzes CIEDE2000 color \cite{ciede2000} difference metric followed by average pooling. It allows us to analyze color distortions introduced by the neural codec. The second method, named ``Texture metric'', computes MS-SSIM values for local regions and compares them with ones compressed by non-neural-network-based JPEG codec. This approach detects distortions of high-frequency textures in compressed images.

\begin{figure}[h]
  \centering
   \includegraphics[width=\linewidth]{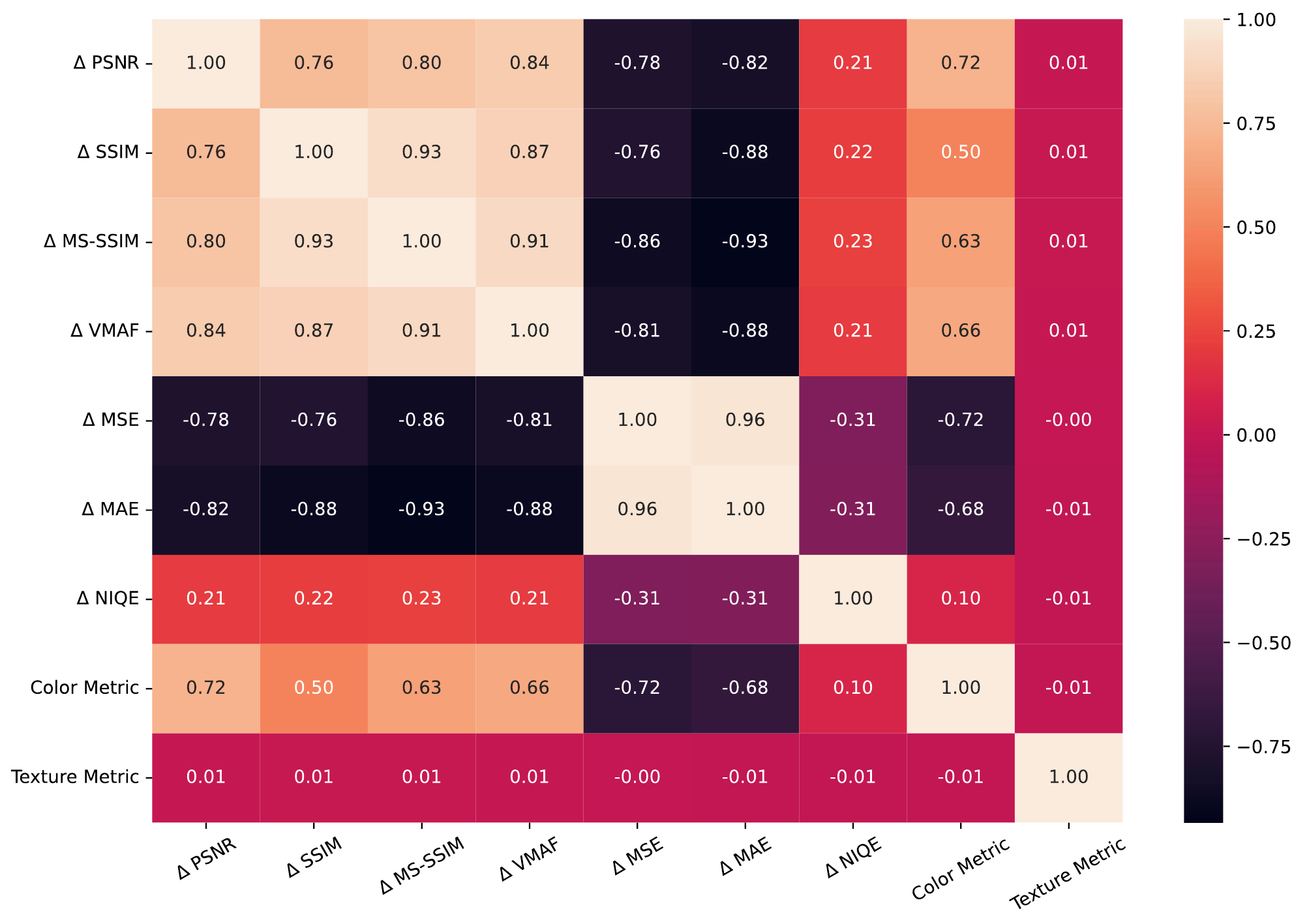}
  \caption{Spearman Correlation Coefficients between different quality scores. Calculated on a subset of $\sim$ 3000 images with compression artifacts found by Color or Texture metric.}
  \label{fig:artifacts_correlations}
\end{figure}

Firstly, we analyze correlations between these two methods for artifact assessment and other metrics introduced earlier. Fig.~\ref{fig:artifacts_correlations} presents a heatmap of pairwise Pearson correlation coefficients. We filter out images where artifacts metrics found no artifacts and measure on $\sim$3000 images. There are strong correlations between $\Delta$-based metrics, except for $\Delta$NIQE. Moreover, the Color metric also shows correlations above 0.5, with most reaching 0.72 correlation with $\Delta$PSNR. However, the Texture metric shows minimal correlation with other metrics, likely due to the smaller resolution of selected datasets, which limits the presence of fine textures. This indicates that reductions in quality metrics (PSNR, SSIM, etc.) are often due to color artifacts induced by adversarial attacks rather than typical compression artifacts.

\begin{figure*}[htb]
  \centering
   \includegraphics[width=0.8\linewidth]{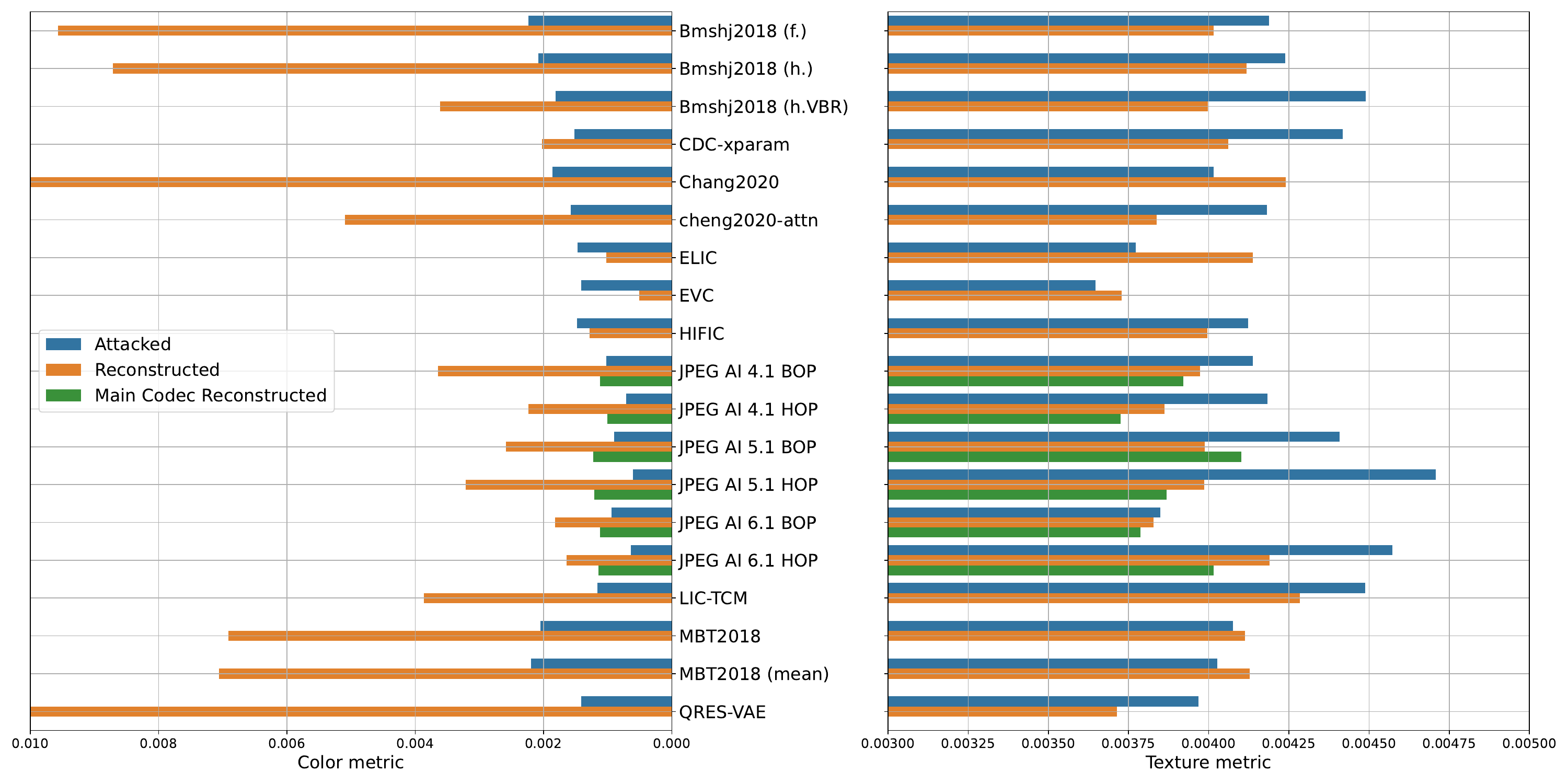}
  \caption{Artifact detecting methods for evaluated codecs. The color metric is on the left plot, and the texture metric is on the right one.}
  \label{fig:artifacts_metric_barplot}
\end{figure*}

Fig.~\ref{fig:artifacts_metric_barplot} shows average values for evaluated artifact metrics split by the codec. It confirms that color artifacts in reconstructed images are stronger than in attacked ones before decompression, making adversarial attacks successful. Cheng2020 and QRES-VAE codecs showed the worst results for the Color metric and average results for the Texture metric. On the other hand, all versions of JPEG AI performed relatively well and introduced the least color artifacts compared to other codecs. The main codec of JPEG AI, at the same time, shows even better results compared to the regular decoder in terms of artifacts. We can also see that with every new version, the amount of artifacts of JPEG AI consistently decreases, showing improving robustness throughout the development of the codec.
Regarding the Texture metric, all the codecs showed comparable results. Notably, artifacts on reconstructed images are often fewer than those of non-decompressed attacked images, indicating that decoders effectively mitigate these artifacts. In this case, versions of JPEG AI have no significant difference.

\subsection{Attacks' transferability}

\begin{figure}[h]
  \centering
\includegraphics[width=\linewidth]{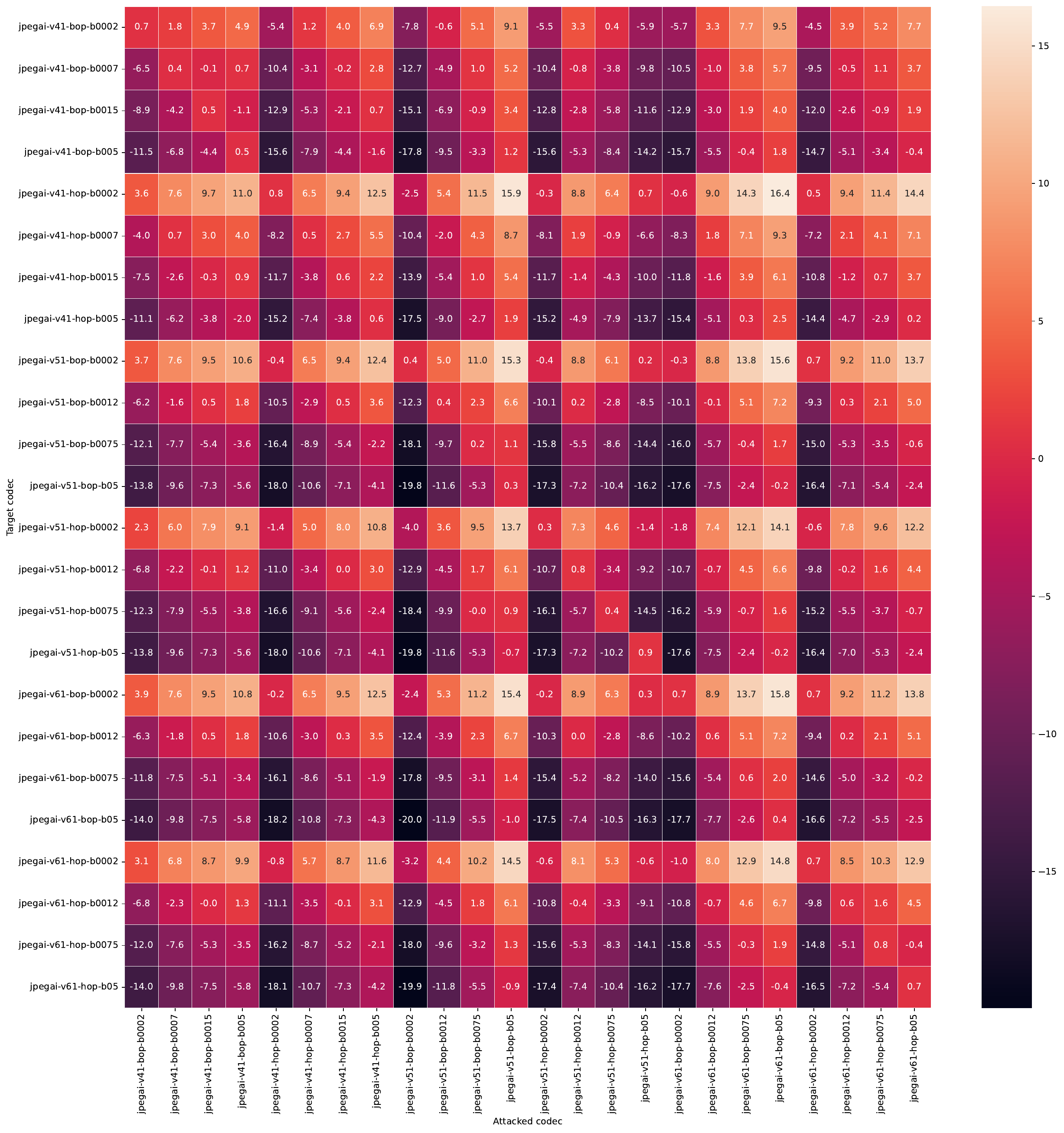}
   \caption{Transferability of adversarial attacks on one JPEG AI version to others. The figure shows $\hat{\Delta}$VMAF described in Equation \ref{eq:delta_transf}.}
\label{fig:jpeg_transferability}
\end{figure}

This section presents the transferability of adversarial examples generated for one NIC to others. We have conducted two experiments: using all NICs presented in our paper and only JPEG AI versions. We compare different bitrates of chosen codecs, so they are included several times. 
The JPEG AI transferability experiment results are in Fig.~\ref{fig:jpeg_transferability}. We ran experiments on all losses for two presets and averaged the results. The results show high transferability between different bitrates of specific versions of JPEG AI, especially from lower bitrates to higher ones. Transferability also exists for other versions of JPEG AI.

\subsection{Defences efficiency}
\begin{figure}[h]
  \centering
   \includegraphics[width=\linewidth]{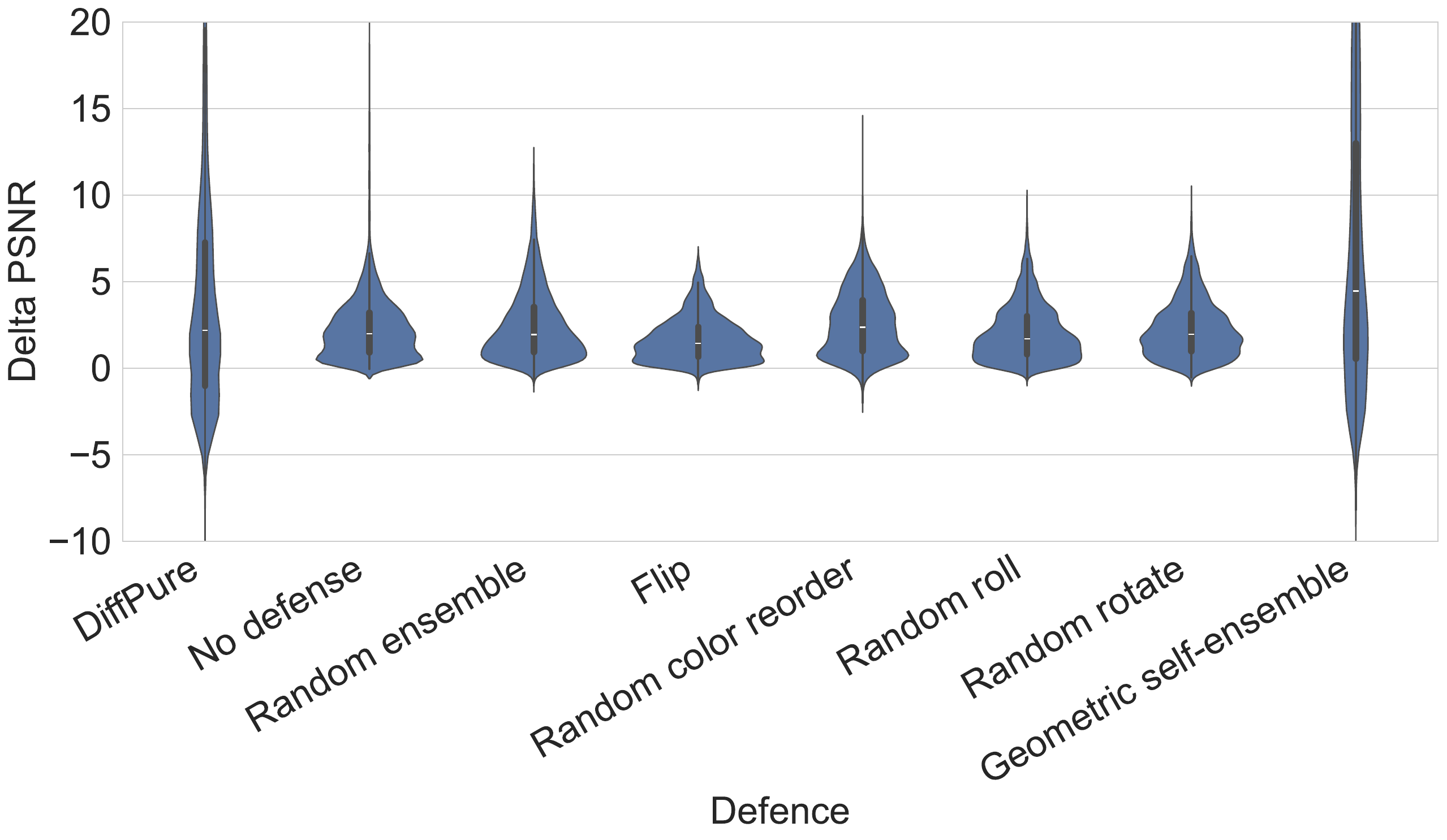}
   \caption{$\Delta \text{PSNR}$ of different defense techniques.}
   \label{fig:defense_perfomance}
\end{figure}

In this subsection, we evaluate defenses on JPEG AI in the way described in section \ref{sec:defense_methodology}. We measure the $\Delta$PSNR change between different defense strategies for FTDA default loss for FTDA, I-FGSM, PGD, and SSAH. The results can be seen in Fig.~\ref{fig:defense_perfomance}. 
Simple reversible defenses can negate adversarial attacks on NIC models. The results demonstrate that Flip, Random Ensemble, and Random Roll are somewhat effective in reducing their effect. 

\subsection{Codecs comparison without attacks}
\begin{figure}[htbp]
  \centering
   \includegraphics[width=\linewidth]{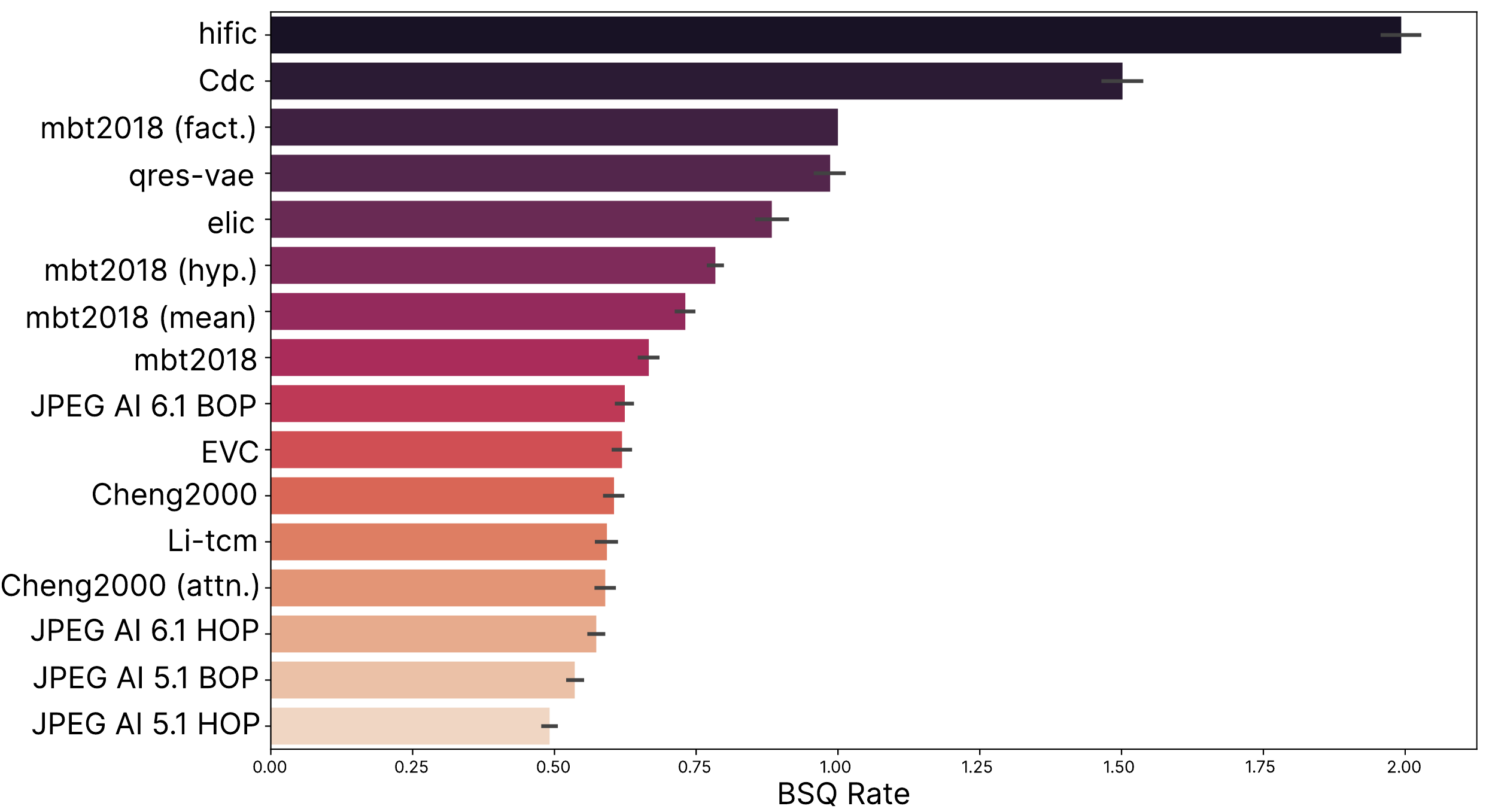}
   \caption{BSQ-rate for NIC methods by VMAF quality metric. Averaged for all tested datasets.}
   \label{fig:barplot-bsqrate}
\end{figure}

Fig.~\ref{fig:barplot-bsqrate} shows BSQ-rate~\cite{zvezdakova2020bsq} (bitrate savings for the same image quality) relative to mbt2018 codec. JPEG AI shows superior compression performance: over 50\% less bitrate with the same quality by VMAF. Surprisingly, Cheng2020 is also in top positions, despite this method is much older than the competitors.

\section{Conclusion}

This study offers an in-depth evaluation of JPEG AI and other neural image codecs (NICs), focusing on compression efficiency, adversarial robustness, and transferability of attack effects. Our findings highlight JPEG AI’s substantial improvements in bitrate efficiency, achieving over 50\% reduction at equivalent VMAF scores compared to mbt2018 codec. This bitrate saving reflects significant progress in compression technologies, positioning JPEG AI as a leading codec in maintaining high image quality with lower bandwidth requirements. While Cheng2020 still shows competitive results, JPEG AI surpasses it in multiple quality metrics.

%TODO надо написать что-то про то какая версия jpeg ai в итоге лучшая

Our study confirms that robustness analysis is a crucial evaluation criterion for NICs models. Although JPEG AI shows high robustness, it is still vulnerable to attacks, and further investigation into specialized adversarial defenses is vital. On the other hand, assessing attack success in NICs remains challenging compared to other computer vision tasks, such as image classifiers. Artifacts that emerge during decoding require specialized models for detecting and assessing, highlighting a relevant area for future research. Additionally, we demonstrate that attacks can transfer to new and potentially proprietary versions of the JPEG AI, and robustness testing becomes essential before widespread distribution and adoption.

% The study’s analysis of adversarial attacks reveals that FTDA losses introduce the greatest visual disturbances, especially in terms of $\Delta L_{\infty}$, highlighting a vulnerability that may be addressed in future NIC developments. Interestingly, reconstruction losses appear to improve quality in decompressed images, pointing to robustness within some NICs under attack scenarios. 

% \begin{figure}[h]
%   \begin{subfigure}{.5\textwidth}
%   \centering
%     \includegraphics[width=\linewidth]{ICLR 2025 Template/imgs/barplot_color.pdf}
%   \end{subfigure}%
%   \begin{subfigure}{.5\textwidth}
%   \centering
%     \includegraphics[width=\linewidth]{ICLR 2025 Template/imgs/barplot_texture.pdf}
%   \end{subfigure}
%   \caption{TODO}
%   \label{fig:artifacts_metric_barplot}
% \end{figure}

{
    \small
    \bibliographystyle{ieeenat_fullname}
    \bibliography{main}
}


\end{document}